\documentclass[sigconf,screen,nonacm]{acmart}
\AtBeginDocument{%
  \providecommand\BibTeX{{%
    \normalfont B\kern-0.5em{\scshape i\kern-0.25em b}\kern-0.8em\TeX}}}

\usepackage{enumitem,amsmath}
\usepackage[symbol]{footmisc}
\usepackage{xcolor}
\usepackage{algpseudocode}
\newlist{todolist}{itemize}{2}
\setlist[todolist]{label=$\square$}
\usepackage{pifont}
\usepackage{courier}

\usepackage[normalem]{ulem} %
\usepackage{natbib}
\usepackage{enumitem}

\begin{document}

\title{JAX-LOB: A GPU-Accelerated limit order book simulator to unlock large scale reinforcement learning for trading}

\author{Sascha Frey\footnotemark[1]\footnotemark[2]}
\affiliation{
    \institution{Department of Computer Science\\University of Oxford}
    \country{UK}
}

\author{Kang Li\footnotemark[1]}
\affiliation{
    \institution{Department of Statistics\\University of Oxford}
    \country{UK}
}

\author{Peer Nagy\footnotemark[1]}
\affiliation{%
  \institution{Oxford-Man Institute of Quantitative Finance\\ University of Oxford}
  \country{UK}
}
\author{Silvia Sapora}
\affiliation{
    \institution{Foerster Lab for AI Research\\University of Oxford}
    \country{UK}
}
\author{Chris Lu}
\affiliation{
    \institution{Foerster Lab for AI Research\\University of Oxford}
    \country{UK}
}

\author{Stefan Zohren}
\affiliation{
    \institution{Man-Group\\Oxford-Man Institute of Quantitative Finance\\University of Oxford}
    \country{UK}
}

\author{Jakob Foerster}
\affiliation{
    \institution{Foerster Lab for AI Research\\University of Oxford}
    \country{UK}
}

\author{Anisoara Calinescu}
\affiliation{
    \institution{Department of Computer Science\\University of Oxford}
    \country{UK}
}

\renewcommand{\shortauthors}{Frey, Li and Nagy, et al.}

\begin{abstract}
Financial exchanges across the world use limit order books (LOBs) to process orders and match trades. For research purposes it is important to have large scale efficient simulators of LOB dynamics. LOB simulators have previously been implemented in the context of agent-based models (ABMs), reinforcement learning (RL) environments, and generative models, processing order flows from historical data sets and hand-crafted agents alike. For many applications, there is a requirement for processing multiple books, either for the calibration of ABMs or for the training of RL agents. We showcase the first GPU-enabled LOB simulator designed to process thousands of books in parallel, with a notably reduced per-message processing time. The implementation of our simulator -- JAX-LOB -- is based on design choices that aim to best exploit the powers of JAX without compromising on the realism of LOB-related mechanisms. We integrate JAX-LOB with other JAX packages, to provide an example of how one may address an optimal execution problem with reinforcement learning, and to share some preliminary results from end-to-end RL training on GPUs. 
\end{abstract}

 

\settopmatter{printfolios=true}
\maketitle
\footnotetext[1]{Equal contribution}
\footnotetext[2]{Correspondance address: sascha.frey@st-hughs.ox.ac.uk}

\section{Introduction}
Markets, i.e. matching buyers and sellers while determining a price, are a crucial component of modern economies. In many instances these markets rely on \textit{auction mechanisms} for their operation. At the most basic level, for a single good sold at a specific time, these mechanisms allow all buyers to state how much they are willing to pay for a given good, and the good then goes to the highest bidder. 

In contrast, in financial markets the price finding process usually needs to happen \textit{continuously} and for \textit{any number} of shares during the operating hours of the market to ensure liquidity. 
To accomplish this, the limit order book (LOB) mechanism is used in most modern electronic exchanges as a price finding tool.

At a high level, LOBs implement an \textit{any-time} auction mechanism that allows all market participants to submit \textit{buy} and \textit{sell} orders specifying a \textit{quantity} of stocks and a \textit{price} at which they are willing to trade. As soon as there are compatible  buy and sell orders, i.e. where the buying price of one order at least matches the selling price of another order, a trade happens.%

Due to their central role in the financial system, the ability to accurately and efficiently model LOB dynamics is extremely valuable. For example, it might allow a financial company to offer better services or may enable the government to predict the impact of financial regulation on the stability of the financial system.

Practically, there are a number of possible scientific approaches for taking advantage of such an LOB simulator, including agent-based models, reinforcement learning (RL), and generative models. 

However, due to the low signal-to-noise ratio a common aspect of these approaches is the need for large scale simulations that take advantage of modern compute hardware and parallelism. Addressing this, \textbf{a GPU accelerated LOB simulator is the core contribution} of our paper.  
One specific task of interest is the execution of orders by brokers or funds, where the goal is to sell/buy a specific quantity of stocks over a given time frame at the best possible price. 
This \textit{optimal execution problem} is both a motivation and an example use-case for the work presented in this paper. 

In Section \ref{sec:background} we provide a brief overview of the LOB and the trade execution problem, and in Section \ref{sec:rel_work} we consider related works both in building LOB simulators and addressing the execution problem with reinforcement learning (RL). In Section \ref{sec:jaxlob_simulator} we present the first GPU-accelerated  LOB simulator using JAX \citep{jax2018github} and discuss the detailed workings of the LOB, both general and specific to our implementation. In Section \ref{sec:gymnax} we lay the groundwork for the use of our simulator for RL, an application domain which benefits significantly from the parallelism enabled by our implementation.

In Section \ref{sec:exec_env} we contribute to solving the optimal execution problem with RL by wrapping the JAX-LOB simulator in a JAX-native \textit{gymnax} \citep{gymnax2022github} execution environment. In our environment, both the experience rollout (i.e. agents interacting with the world to collect data) and the learning updates (i.e. agents training on the data) are conducted on the same GPU which avoids the GPU-CPU communication bottleneck. Finally, in Section \ref{sec:learningtask} we integrate the environment with \textit{PureJaxRL} \citep{lu2022discovered} and use recurrent proximal policy optimization (PPO) to train an execution agent as an example application. Lastly, we compare the computational speed with equivalent CPU based implementations and show that the maximum speedup occurs when applying JAX-LOB to an RL task, taking full advantage of the parallelism.

Our main contributions are as follows: \\
\textbf{1)} The first GPU-accelerated LOB simulator, JAX-LOB
\\
     \textbf{2)} At least 7x speedup in training an RL agent, compared to equivalent CPU-based LOB simulators: 550 versus 74 steps per second during training, on the same hardware (Section \ref{sec:training_speedup}).

We commit to open-sourcing JAX-LOB incl. the RL integration and strongly believe it will unlock a number of new research avenues for the community.

\vspace{-2mm}

\section{Background}
\label{sec:background}
\textbf{The LOB}, the underlying data structure and associated mechanisms which power modern electronic exchanges, has been studied extensively \citep{gould_limit_2013,bouchaud_trades_2018}. The LOB is the collection of all orders to buy or sell a security that have yet to be matched. Two types of orders may be submitted to the LOB at any time: limit orders which require the specification of a price and a quantity, and market orders, which require only a quantity. Traders are free to submit orders at any time, at which point they are either matching with a compatible opposite order in the book, or added to the book -- in the case of a limit order. Further details of the LOB are discussed in Section \ref{sec:jaxlob_simulator}, together with details specific to our implementation. LOB simulators have a number of application domains, including generative models \citep{nagy2023generative}; in this paper we focus on their use to train trading agents by using RL for trading tasks such as market making \citep{beysolow2019market,ganesh2019reinforcement} or trade execution \citep{nevmyvaka2006reinforcement}.

\textbf{JAX} \citep{jax2018github} is an accelerator agnostic framework that enables just-in-time (JIT) compilation using accelerated linear algebra (XLA) \citep{Google_7AD}, automatic differentiation, and automatic vectorization which can easily be executed on the GPU. The framework is designed for high-performance machine learning research and forms the basis of both the \textit{gymnax} \citep{gymnax2022github} and \textit{PureJaxRL} \citep{lu2022discovered} frameworks which we leverage, using optimal trade execution as an example problem. JAX has also been used for other applications where massive parallelism is advantageous, such as fluid dynamics \citep{kochkov_machine_2021} or molecular dynamics \citep{jaxmd2020}. Whilst the focus of this paper is on the JIT compilation and GPU vectorization, the differentiation of LOB models for calibration \citep{quera-bofarull_challenges_2023} is a topic that is thus far underexplored in the literature.

\textbf{Optimal trade execution} is a well-studied problem in finance, with traditional approaches using methods from stochastic optimal control and analytic market-impact functions  \citep{bertsimas_optimal_1998,almgren_optimal_2001,kyle_continuous_1985,obizhaeva_optimal_2013} to derive optimal policies. The aim of optimal trade execution is to buy or sell a specified number of stocks in a specified time frame, obtaining the best possible average price at low risk. 
This usually requires minimizing the cost due to market impact, especially when the trades are large compared to the liquidity of the market \citep{almgren_optimal_2001}.

\section{Related work}
\label{sec:rel_work}
There are a number of implementations that reconstruct the mechanics of the LOB \citep{byrd_abides_2020,belcak_fast_2022,jerome_market_2022,paulin_understanding_2019_custom,julialob2020github}. The implementations may be further subdivided by the application they are used for: agent-based models (ABMs) and market replay. Frameworks built for ABMs, such as the agent-based interactive discrete event simulation (ABIDES) framework \citep{byrd_abides_2020} or the multi-agent exchange environment (MAXE) \citep{belcak_fast_2022} implement a heterogeneous set of agents to submit fictional orders to the order book. ABIDES is implemented in standard Python and has an extension, ABIDES-gym \citep{amrouni_abides-gym_2022}, which creates a gym-like interface for reinforcement learning. MAXE is implemented in C++ and benefits from increased speed due to compilation \citep{belcak_fast_2022}. In market replay models, such as the reinforcement learning for market making (RL4MM) framework \citep{jerome_market_2022}, historical orders based on message data, such as the ``limit order book system - the efficient reconstructor'' (LOBSTER) \citep{huang_lobster_2011} data set are submitted. The primary limitation of ABMs is the difficulty in creating sufficiently realistic agent behaviors. Despite significant recent efforts \citep{vyetrenko_get_2020}, this remains a limitation, especially when compared to replayed market data, which directly replicates the data distribution. However, the strength of ABMs is also a key limitation of market replay \citep{jerome_model-based_2022}. The implementation of strategic agents results in a dynamically reactive order flow based on outside intervention (e.g. a RL agent), whereas replayed historical order flow will remain unchanged. JAX-LOB, while originally designed to process data from LOBSTER \citep{huang_lobster_2011}, may process messages from any source, including strategic agents. In contrast to the discussed implementations, JAX-LOB runs on the GPU, allowing for large-scale parallelization. 

\citet{ddqn_opt_ex_ning18} use double deep Q Networks (DDQN) \citep{dddqn_hasselt2016} to solve the optimal execution problem using only market orders. These are, however, generally sub-optimal as trades may ``walk the book'', i.e. consume the entire available volume on one or more levels of the LOB, and eliminate potential gains from the spread by using passive limit orders. Furthermore, market impact is stylized by using a penalty function for larger orders. In this setting, they outperform the time-weighted average price (TWAP) benchmark in 7 out of 9 evaluated stocks. 
\citet{daberius19} compares the use of DDQN \citep{dddqn_hasselt2016} and PPO \citep{ppo_schulman17} for optimal execution using only market orders in a stochastic model of price dynamics without price impact. In building a LOB simulator, we provide the tools to move away from having to stylize the market impact and create an environment that allows for the submission of limit orders at various prices, thus avoiding the risk of inadvertently consuming volume on multiple levels of the book. Nevertheless, the exclusive use of historical LOB data means that only direct market impact is accounted for. Such historical messages have no strategic behavior that may react to an RL agent's actions. 

Addressing this drawback, \citet{karpe20} uses DDQN in an agent-based simulation environment based on the ABIDES simulator \citep{byrd_abides_2020}, which replays real LOB message data while adding additional heuristic agents, which react to changes in the LOB state caused by the RL agent. Their simulation setup allows addressing optimal execution and optimal level placement jointly. They find that their agent converges to a TWAP strategy. While additional agents allow for more realistic market impact in principle, the ABIDES heuristics use a somewhat simplistic momentum strategy based on aggregate summary statistics, instead of a learned rational strategy. Similarly, \citet{ma_opt_ex_ppo_fang22} use PPO to address the combined optimal execution and placement problem using the approach of replaying historical data in the ABIDES \citep{byrd_abides_2020} simulator. Their agent learns placement of up to three levels of the LOB concurrently. They use a large number of engineered features, including technical indicators, an attention-based network architecture, dubbed Dual-Window Denoise PPO, and a reward formulation based on execution cost and an imitation term, steering the policy towards a TWAP policy. Results indicate a potential improvement over TWAP but suffer from high standard errors. The environment we set up aims to move away from such engineered feature spaces and reward functions, and focuses on using recurrent neural networks exclusively to automatically extract features and account for the longer-term memory of the agent.

\vspace{-1mm}

\section{The JAX-LOB simulator}
\label{sec:jaxlob_simulator}
Some of the key challenges in applying deep reinforcement learning to trade execution and other high-frequency tasks are the low signal-to-noise ratio, the risk of overfitting to specific training days, and building a simulator with realistic market impact. A straightforward way to compensate for the first two issues is to increase the number of state-action transitions available for training. To speed up the generation of observations of faithful LOB representations using high-frequency data we use JAX \citep{jax2018github}.

We make a number of design decisions to address some of the constraints of JIT-compiled JAX code: 
\begin{itemize}
    \item Pure functions without side effects (e.g. global variables)
    \item Fixed array size and type
    \item Control-flow that can be compiled and parallelized efficiently (See Section \ref{sec:vmap}).
\end{itemize}

Most CPU implementations of LOBs are based on hash-maps, queues, doubly-linked lists and sorted dictionaries \citep{belcak_fast_2022,byrd_abides_2020,jerome_market_2022} which enable quick access to data and maintain sorting of the book throughout. Given the JAX requirement of compile-time fixed-size arrays, implementing a similar structure means memory must be pre-allocated for all price-levels and orders. The use of arrays means that re-ordering upon removal of entries is far more costly than with linked lists. We, therefore, opt for an architecture that does not use a tree-like structure, nor do we keep orders sorted at all times. Instead, we define two arrays $\mathcal{A}$ and $\mathcal{B}$ to represent the sides of the order book containing all active ask and bid orders, respectively.
\begin{equation}
        \mathcal{A} = [\boldsymbol{a_1}, ..., \boldsymbol{a_N}]  \quad \mathcal{B}= [\boldsymbol{b_1}, ..., \boldsymbol{b_N}]\\
\end{equation}
\begin{equation}
\label{eq:features}
        \boldsymbol{o_i}=[P_i, Q_i, OID_i, TID_i, Ts_i, Tns_i]  \in \mathcal{A} \cup \mathcal{B}, \quad i \in [1,N]
\end{equation}
Each side of the book has a fixed capacity for $N$ orders, where each order $\boldsymbol{o}$ has six features \eqref{eq:features}: Price $P$, Quantity $Q$, Order ID $OID$, Trader ID $TID$, Time (seconds) $Ts$, and Time (nanoseconds) $Tns$. Empty positions in $\mathcal{A}$ or $\textbf{B}$ are indicated by setting all features to -1. The size $N$ must be chosen such that the book is not saturated during a given experiment.

\subsection{Basic operations}
\label{sec:basicops}
There are three operations that can be applied to either side of the order book; the \textit{addition} of a new order, the \textit{cancellation} of an existing order, and the \textit{matching} of an existing order with an incoming order on the other side of the book followed by its removal from the book.

\begin{enumerate}
\item The addition of an order requires the identification of an empty position ($\boldsymbol{o_i}=\boldsymbol{-1}$) in the array, and the insertion of the order-specific data into the correct fields. 

\item Cancellation requires locating the order ID ($OID$) of the order to be canceled and removing the corresponding quantity from the book.  

\item During the matching operation, an incoming aggressive order, denoted $\boldsymbol{o_a}$, is matched against an existing order on the other side of the book, the standing order $\boldsymbol{o_s}$. The matching logic aims to find the remaining quantities for both the aggressive ($Q_a\in \boldsymbol{o_a}$) and standing ($Q_s\in \boldsymbol{o_s}$) orders using the following operations
\begin{align*}
\label{eq:test}
Q_s'&=max(0,Q_s-Q_a),\quad Q_a'=Q_a-Q_s.
\end{align*}
When two orders are matched, a trade $\boldsymbol{t_j}$ is logged \eqref{eq:trade}. 
\begin{equation}
\label{eq:trade}
        \boldsymbol{t_j}=[P_j, Q_j, OID_{a,j}, OID_{s,j}, Ts_j, Tns_j] 
\end{equation}
Whereby:
\begin{align*}
P_j&=P_s\\
Q_j&=Q_s-Q_s'\\
OID_{a,j}&=OID_a \in \boldsymbol{o_a}\\
OID_{s,j}&=OID_s \in \boldsymbol{o_s}\\
Ts_j, Tns_j &= Ts_a, Tns_a
\end{align*}
Up to N trades are logged in a fixed-size array $\mathcal{T}$ due to constraints of the XLA compiler \eqref{eq:trades}. 
\begin{equation}
\label{eq:trades}
        \mathcal{T} = [\boldsymbol{t_1}, ..., \boldsymbol{t_N}]^T
\end{equation}
Again, \textbf{$\boldsymbol{t_j}=\boldsymbol{-1}$} indicates an empty slot.
\end{enumerate}
For all operations, upon completion, both $Asks$ and $Bids$ are checked for orders $\boldsymbol{o_i}$ where $Q_i \leq 0$, in which case we set $\boldsymbol{o_i} = \boldsymbol{-1}$.

A single incoming order may match with more than just one standing order. The matching logic contains a while loop that repeatedly attempts to match the incoming order against the next-best standing order on the opposite side of the book. The best standing order $Best(\boldsymbol{o_s})$ is defined by the price-time order priority algorithm, the most commonly used LOB matching algorithm \citep{gould_limit_2013}. The best order is the one with the ask price or bid price \eqref{eq:best_price}. If multiple orders share this price, the one with the earliest arrival time is considered.  
\begin{equation}
\label{eq:best_price}
        P_{ask} = \underset{\boldsymbol{o} \in \mathcal{A}}{min}(P_i) \quad P_{bid} = \underset{\boldsymbol{o} \in \mathcal{B}}{min}(P_i)
\end{equation}

The while-loop continues as long as the book is non-empty, $Q_a >0$, and the prices overlap as follows
\begin{align*}
    P_a &\leq P_s &\text{if marketable sell order},\\
    P_a &\geq P_s &\text{if marketable buy order}.
\end{align*}

Based on the above descriptions it is intuitive to expect varying computational complexity across these basic operations. To validate this intuition, we time the basic operations in order books of different maximum capacities $N$ (Table \ref{tab:simpleopstimes}). The operations take longer with increasing capacity in the book $N$, and as expected, the slowest operation (\textit{Match}) takes more than twice as long to execute as the fastest (\textit{Cancel}).
\begin{table}[h]
\centering
\vspace{-3mm}
\caption{Average run-time for the three basic operations that may be applied to the LOB for different sizes $N$. The tests are conducted on an Nvidia 2080 Ti and averaged across 1000 serial trials. The book is initially filled to one-third capacity with random orders.}
\label{tab:simpleopstimes}
\vspace{-3mm}
\begin{tabular}{ |c|c|c|c| } 
 \hline
 Capacity $N$& Add (ms) & Cancel (ms) & Match (ms)\\ 
 \hline
 10 & 0.115 & 0.081 & 0.184\\ 
 100 & 0.117 & 0.095 & 0.206\\ 
 1000 & 0.162 & 0.093 &0.243\\ 
 \hline
\end{tabular}
\vspace{-3mm}

\end{table}

The relative size of the incoming order $Q_a \in \boldsymbol{o_a}$ will have a significant impact on the computation time for the matching operation. All else being equal, a larger quantity requires the consideration of more standing orders to match the arriving order, thus computing more iterations of the matching logic. A simple way to illustrate this is to consider an order book with capacity $N=100$, filled to one-third capacity, and to submit market orders of varying sizes (Table \ref{tab:matchvsquant}). This increase in time is not drastic until particularly large orders are submitted, but shows the importance of constraining the action space to reasonable quantities to ensure quick execution. 
\vspace{-2mm}

\begin{table}[h]
\centering
\caption{Analysis of the effect of incoming order size on the time required to complete the matching operation. As the size increases, more standing orders need to be considered to fully match the incoming order. The capacity for the book is set to $N=100$. Testing is done on an Nvidia 2080 Ti GPU.}
\label{tab:matchvsquant}
\vspace{-2mm}
\begin{tabular}{ |c|c| } 
 \hline
 Market $Q_a$ & Time to match (ms) \\ 
 \hline
 0 & 0.132 \\ 
 10 & 0.206 \\ 
 500 & 0.271 \\ 
 1,000 & 0.336 \\ 
 10,000 & 2.326 \\ 
 \hline
\end{tabular}

\vspace{-7mm}

\end{table}

\subsection{Message types}
Which of the three basic operations (Section \ref{sec:basicops}) is called depends on the type $T$ of message $\boldsymbol{m}$ \eqref{eq:message} passed to the order book.
\begin{equation}
\label{eq:message}
    \boldsymbol{m}=[T ,S , Q , P, OID, TID, Ts, Tns]
\end{equation}

The structure of data contained in messages $\boldsymbol{m}$ is:
\begin{enumerate}
    \item $T$ - The type of message: limit orders ($T=1$), cancel orders ($T=2$), delete orders ($T=3$), and market orders ($T=4$).
    \item $S$ - The side of the order: bid ($S=1$) or ask ($S=-1$).
    \item $P$ - The price at which the order should be submitted or the price of the order to be cancelled/deleted. This is ignored for market orders.
    \item $Q$ - The size of the order, i.e. the quantity to buy/sell or the to remove from an existing order for cancel/delete orders. 
    \item $OID$ - The unique order ID. 
    \item $TID$ - The trader ID, an identification marking the source of the order. Used to identify orders from individual agents. 
    \item $Ts, Tns$ - The time of receipt of the order, split into two fields representing the time as full seconds and fractional digits as nanoseconds represented by 32-bit integers. 
\end{enumerate}

While the reader familiar with the LOBSTER \citep{huang_lobster_2011} data set will recognize significant similarities to the message structure in this data set, they are not identical. We take the following steps to process orders of different types:

\begin{enumerate}
    \item Limit orders are first matched against the opposite side of the book, the remaining quantity is then added as a single new order to the corresponding side.
    \item Cancel and delete orders are treated identically, by calling a cancellation (Section \ref{sec:basicops}).
    \item Market orders are matched against the opposite side with a price $P_m=0$ if an ask order, or $P_m=max\_int$ if a bid order. Any unmatched quantity is disregarded. 

\end{enumerate}

Considering each type and side as eight separate cases allows defining explicit functions for each case, enabling the use of a single conditional statement upon receipt of the message, rather than multiple branches within the matching logic, which we find improves performance under \texttt{vmap} (Section \ref{sec:vmap}). The compute time (Table \ref{tab:msgtimes}) to process each of the three message types varies and is related to the basic operations required. When compared to the results in Table \ref{tab:simpleopstimes} the time is only marginally increased by the branching statement, and there is still a notable difference across different order types. 
\vspace{-3mm}
\begin{table}[h]
\centering
\caption{Time to process different message types. The last column differs from the second column in that the price crosses the spread, requiring a matching operation. Tests are done on an Nvidia 2080 Ti GPU, the book capacity is $N=100$.}
\label{tab:msgtimes}
\vspace{-3mm}
\begin{tabular}{ |p{1.25cm}|p{1.5cm}|p{1.6cm}|p{2.4cm}| } 
 \hline
 Capacity $N$ & Limit Order (ms) & Cancel Order (ms) & Limit Order across book (ms) \\ 
 \hline
 10 & 0.108 & 0.077 & 0.163\\ 
 100 & 0.157 & 0.097 & 0.203\\ 
 1000 & 0.195 & 0.095 &0.247\\ 
 \hline
\end{tabular}

\vspace{-5mm}

\end{table}

\subsection{Vectorising map - \texttt{vmap}}
\label{sec:vmap}

The benefit of implementing the LOB in JAX is derived from parallelizing computation on the GPU. However, the processing of messages sent to the order book in a continuous double-auction is intrinsically a serial process: both the treatment of messages and the internal matching logic must be strictly ordered. This is in opposition to a classical auction with, say, a central clearing process \citep{scholl_how_2021}, or even stochastic models of the limit order book \citep{cont_stochastic_2010}. Therefore, to achieve parallelism we process multiple books in parallel using the \texttt{vmap} operator.

One particularity of the \texttt{vmap} operator is that it transforms a number of control-flow statements into \texttt{select} statements which execute all of the conditional branches at run time. In the case of limit order books, this means that each of the possible eight cases is computed and the runtime is therefore limited by the slowest branch. This can be observed when looking at the timed results (Table \ref{tab:msgtimesvmap}) when processing the same message across $N\_books=1000$ identical order books in parallel. The time no longer varies across different types of messages but only depends on the capacity. The reason we avoid continuous sorting is that when removing orders in an array-based view, the entire array needs to be re-sorted. This order sorting is comparatively costly, which would not be problematic if called rarely. However, the aforementioned behavior of the \texttt{vmap} operator means that it would be called for every message, even if no order needs to be removed. Even though we have done extensive tests, it is possible that there is still a bottleneck of this nature which would explain the worst-case message processing time of over 2 seconds in Table \ref{tab:msgtimesvmap}.
\begin{table}[h]
\vspace{-2mm}
\centering
\caption{Time to process different message types with \texttt{vmap}. The difference to table \ref{tab:msgtimes} is that messages are processed by $N\_Book = 1000$ books in parallel. Times measure the processing of a message in all books. The effective processing time for a single message can therefore be interpreted as being in $\mu s$ and compared directly to the times in Table \ref{tab:matchvsquant}.}
\label{tab:msgtimesvmap}
\vspace{-3mm}
\begin{tabular}{ |p{1.25cm}|p{1.5cm}|p{1.6cm}|p{2.4cm}| } 
 \hline
 Capacity & Limit Order (ms) & Cancel Order (ms) & Limit Order across book (ms) \\ 
 \hline
 10 & 1.402 & 1.407 & 1.393\\ 
 100 & 2.649 & 2.614 & 2.612\\ 
 1000 & 11.75 & 11.43 &11.47\\ 
 \hline
\end{tabular}
\vspace{-2mm}

\end{table}

The resulting times per message type are compared to two CPU implementations (Table \ref{tab:cputimesmessages}): one implementation of the LOB using RB trees and linked lists as well as a similar implementation named RL4MM \citep{jerome_market_2022} using numpy. The CPU environments are described in more detail in sections \ref{sec:gymnax} and \ref{sec:learningtask}. When parallelizing to more than 1000 books, our JAX-LOB has a faster per-message processing time.

\begin{table}[h]
\centering
\caption{Processing times for messages on the CPU order book implementation based on \textit{linked list} and \textit{RB trees}, and the implementation used in RL4MM \citep{jerome_market_2022} based on \textit{numpy} arrays, and for the JAX-LOB implementation based on \textit{jax arrays} with capacity $N=100$ and $N\_Books = 1000$ books in parallel. A key difference between the CPU implementation and that in RL4MM is the fact that data is pre-loaded, as opposed to being accessed from a database. Both CPU implementations are tested on a single core of an Apple M1 chip and the JAX-LOB on an Nvidia 2080 Ti.}
\label{tab:cputimesmessages}
\vspace{-2mm}
\begin{tabular}{ |p{2cm}|p{1.5cm}|p{1.5cm}|p{1.5cm}| } 
 \hline
 Order type & RL4MM Time ($\mu s$ )& CPU \quad Time ($\mu s$ )& JAX-LOB Time ($\mu s$ )\\ 
 \hline
 Limit & 4.91  &5.3&2.6\\ 
 Cancel & 16.12 &3.6&2.6\\ 
 Limit Cross & 37.94 &7&2.6\\ 

 \hline
\end{tabular}

\end{table}

\section{\textit{Gymnax} environments for the LOB}
\label{sec:gymnax}

Our JAX-LOB is well suited for the \textit{gymnax} reinforcement learning environment framework designed around JAX to produce parallel environment roll-outs on the GPU. In the following sections we briefly describe the construction of a general base environment that can be used to derive environments for various tasks, such as market-making, trade execution, or other intraday tasks. We further describe our example order execution environment, which we use in Section \ref{sec:learningtask} to begin training a recurrent PPO-based RL agent. 

\subsection{Base environment}
\label{sec:base_env}
The primary roles of the base environment are to load LOBSTER \citep{huang_lobster_2011} data for replay, provide an interface to the JAX order book, and set up a skeleton trading environment based on JAX-LOB. 

\subsubsection{Data loading}
\label{sec:data_load}
\begin{figure}[h]
  \centering
  \includegraphics[width=0.9\linewidth]{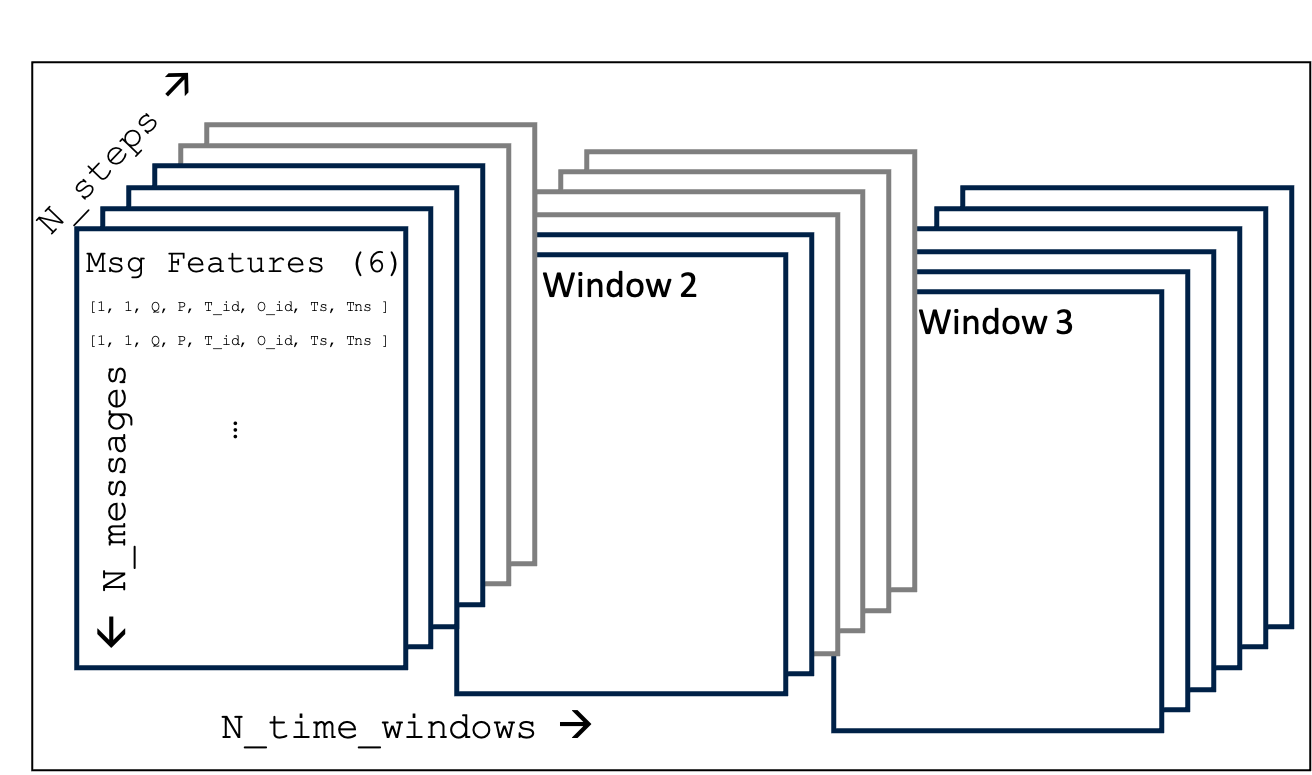}
  \caption{Structure of data loaded into the \textit{gymnax} environments. In our experiments, each step processes $\texttt{N\_messages}=100$ messages. The number of $\texttt{N\_steps}$ is variable with a fixed time of 30 minutes but the data is padded with zeros (grey boxes). There are $\texttt{N\_time\_windows}$ per trading day of data.}
  \label{fig:datawindows}
\vspace{-6mm}

\end{figure}

We opt to pre-load message data in fixed-size, non-overlapping time windows (figure \ref{fig:datawindows}) to save time during execution. The number of messages per step is constant, with a variable time duration per step. Conversely, the duration of a single time window is fixed, implying a variable horizon \citep{amrouni_abides-gym_2022,jerome_market_2022}, i.e. maximum number of steps. This allows the execution of trades with a strict time constraint. To satisfy the fixed-size array constraints, additional steps (indicated by the gray borders in Figure \ref{fig:datawindows}) with all values set to zero are added. These steps are never processed as the episode termination condition is satisfied beforehand.

Further, the order book state (Level-2) of the first ten levels of the book at the beginning of each time window is used to initialize the book. Unlike the message data, this does not provide insight into whether price levels contain one or multiple orders, nor does it provide any order IDs. We work around this by assuming that there is exactly one order per price level containing the sum of the listed volumes. We use $OID_i \in [-inf,-9000]$ starting at $-9000$ and descending for these initial orders. We allow for the cancellation of these orders if a cancellation message satisfies $P_i = P_{cancel} \text{ and }  OID_i < -9000$.
\vspace{-2mm}

\subsubsection{Core state-space requirements}
\label{sec:core_state}

The base environment defines the core features required in the state space, additional elements may be added for derived environments:
\begin{itemize}
    \item Order book:
    \begin{itemize}
        \item Ask side: Array($N_{orders} \times N_{features}$)
        \item Bid side: Array($N_{orders} \times N_{features}$)
    \end{itemize}
    \item Trades: Array($N_{trades} \times N_{features}$)
    \item Initial time: $[T_s, T_{ns}]$
    \item Current time: $[T_s, T_{ns}]$
    \item ID counter
    \item Window Index 
    \item Step counter 
\end{itemize}

The ID counter is used to generate orders submitted by the agent(s), and is incremented accordingly during an episode.

\subsubsection{Functionality}
\label{sec:base_func}

We keep the description of the base environment brief, leaving consideration of the action and observation spaces, and the reward function to the execution environment in Section \ref{sec:exec_env}. The core functionality of the environment during a step is as follows: 

\begin{enumerate}
    \item The received actions (environment dependent) are transformed into messages to be processed by the JAX-LOB.
    \item The data messages are obtained from the loaded data in the environment parameters using the window index and step counter variables (Section \ref{sec:data_load}) for indexing. 
    \item The current time is updated based on the last data message.
    \item All messages are processed by the JAX-LOB.
\end{enumerate}

The episode terminates when the elapsed time is longer than the episode time.

\subsection{Execution environment}
\label{sec:exec_env}

The execution environment extends the state space and defines the observation and action space, reward function, and a new termination condition. 

\subsubsection{Augmented state space}
\label{sec:exec_state}

The state space in \ref{sec:core_state} is augmented with intra-step data and agent-specific information:
\begin{itemize}
    \item Level 1 data for every processed message from data since the last step:
    \begin{itemize}
        \item Best ask price: Array($N_{messages}$)
        \item Best ask volume: Array($N_{messages}$)
        \item Best bid price: Array($N_{messages}$)
        \item Best bid volume: Array($N_{messages}$)
    \end{itemize}
\item the initial mid-price $\frac{P_{ask}+P_{bid}}{2}$ at the beginning of the episode
\item The size of the execution task 
\item The quantity executed thus far 
\item The total revenue due to execution thus-far
\end{itemize}

\subsubsection{Observation space}
\label{sec:exec_obs}

The state space is internal to the environment. The observation space is accessible by the RL agent and is designed to contain information that is accessible in real-life and avoids redundancy. The structure is a single one-dimensional array of size $N_{messages}\times6 + 2 \times 2 + 4$ containing the following metrics:
\begin{itemize}
    \item Best bid prices between steps: Array($N_{messages}$)
    \item Best ask prices between steps: Array($N_{messages}$)
    \item Mid-prices between steps: Array($N_{messages}$)
    \item The prices $n$ ticks deep on the side of the book of the task between steps
    \begin{equation}
    P_{passive}= 
    \left\{
        \begin{array}{lr}
            P_{ask}+n\times ticksize, & \text{if } task \text{ is } sell \\
            P_{bid}-n\times ticksize, & \text{if } task \text{ is } buy 
        \end{array}
    \right\}
    \end{equation}
    \item The spreads between steps: $spread=P_{ask}-P_{bid}$: \\ Array($N_{messages}$)
    \item The current time: $[T_s, T_{ns}]$
    \item The elapsed time in the episode: $[T_s, T_{ns}]$
    \item The initial mid price at the episode start: $P_{init}$
    \item The price drift: $P_{mid}-P_{init}$
    \item The size of the execution task: $Q_{task}$
    \item The quantity executed thus far: $Q_{exec}$
    \item The Level 1 imbalances between steps: $Q_{ask}-Q_{bid}$ :\\ Array($N_{messages}$)
\end{itemize}

\subsubsection{Action space}
\label{sec:exec_action}
In order to have a rich set of actions whilst maintaining a reasonably simple learning task, we opt for a continuous action space of four dimensions. Similar to \cite{ma_opt_ex_ppo_fang22}, at every step, the agent must choose the size of the order to place at four pre-defined prices:
\begin{enumerate}
    \item The ``far touch'' price: the best price on the opposite side of the book 
    \begin{equation}
    P_{far\_touch}= 
    \left\{
        \begin{array}{lr}
            P_{bid} & \text{if } task \text{ is } sell \\
            P_{ask} & \text{if } task \text{ is } buy 
        \end{array}
    \right\}.
    \end{equation}
    \item The Mid-price (Section \ref{sec:exec_obs}).
    \item The ``near touch'' price: best price on the side of the book of the task. 
    \item The ``passive'' price, as defined in Section \ref{sec:exec_obs}.
\end{enumerate}

\subsubsection{Reward function}
\label{sec:exec_reward}

We propose a reward function that combines the agent's ability to outperform the baseline volume-weighted average price (VWAP) strategy and the ability to use predicted price trends. It is composed of two parts, the advantage and the drift. The former represents the advantage gained over the average price of execution, weighted by volume, during the step. The latter represents the effect of price movements of the average traded price with respect to the mid-price at the beginning of the episode. The parameter $\lambda$ can be adjusted to weigh the effect of this drift. In equations \eqref{eq:rewardfunc} and \eqref{eq:vwap}, the indices $i$ and $j$ respectively refer to the set of executed trades during a step, and the set of executed trades of the RL agent. 
\begin{equation}
\label{eq:rewardfunc}
    R=\sum_{j} Q_j(P_j-P_{VWAP}) + \lambda \sum_j Q_j(P_{VWAP}-P_{init})
\end{equation}
\begin{equation}
\label{eq:vwap}
    P_{VWAP}=\frac{\sum_{i} Q_iP_i}{\sum_{i} Q_i}, \quad \sum_{j} Q_j< \text{ task size }
\end{equation}

\subsubsection{Termination condition and market order submission}
\label{sec:exec_done}

In comparison to the base environment, the execution environment allows for an additional termination condition: the completion of the task, i.e. the execution of the desired quantity. In practice, the step function is modified such that this becomes the only termination condition, by submitting a market order for the remaining quantity one minute before the episode time ends.

\subsection{Rollout cost comparison}
\label{sec:exec_costs}

Before considering the RL problem in Section \ref{sec:learningtask}, we consider the run-time improvements of the execution environment based on the JAX-LOB over comparable gym environments running on the CPU. 

We compare our execution environment with the RL4MM \citep{jerome_market_2022} environment designed for market making, but with comparable core logic. All orderbooks are run with data from January 2nd 2015 for ten levels for the stock TSLA. In the CPU and JAX-LOB environments, 100 messages are processed per step. RL4MM measures step duration in time, but we find that 100 messages are processed roughly every 2 minutes. 

 In the case of the JAX-LOB environment, we run different numbers of environments in parallel on an Nvidia 2080 Ti GPU, whilst the RL4MM and the CPU environments are run in series on an Apple M1 CPU. The improvement (Table \ref{tab:stepstiming}) is roughly a factor 10 over the RL4MM environment and a factor 5 over our CPU implementation for 1000 parallel environments. Increasing the number of environments further is a feasible possibility, especially on GPUs with larger memory, but such economies of scale are eventually limited due to memory. 

\begin{table}[h]
\centering
\caption{Average time per step comparison between the RL4MM environment, our CPU-based \textit{gym} environment, and the JAX-LOB based \textit{gymnax} environment for a varying number of environments in parallel. The CPU environments are run on an Apple M1 processor and the \textit{gymnax} environment is run on an Nvidia 2080 Ti.}
\label{tab:stepstiming}
\vspace{-2mm}

\begin{tabular}{ |c|c|c|c|c|c| } 
 \hline
 - & RL4MM &CPU & \multicolumn{3}{c|}{\textit{gymnax}} \\ 
 \hline
 N\_Envs & -&- & 10&1000&10000\\ 
 Total time (s) & 29.6&33.8 & 69&329&2306\\ 
 N\_steps & 5k &9k& 5k&500k&5000k\\ 
 Time/step (ms) & 5.9 &3.5& 13.8&0.66&0.46\\ 
 \hline
\end{tabular}

\vspace{-2mm}
\end{table}

\section{Using the \textit{gymnax} environment for reinforcement learning}
\label{sec:learningtask}

\subsection{Training loop}
\begin{figure}[h]
  \centering
  \includegraphics[width=\linewidth]{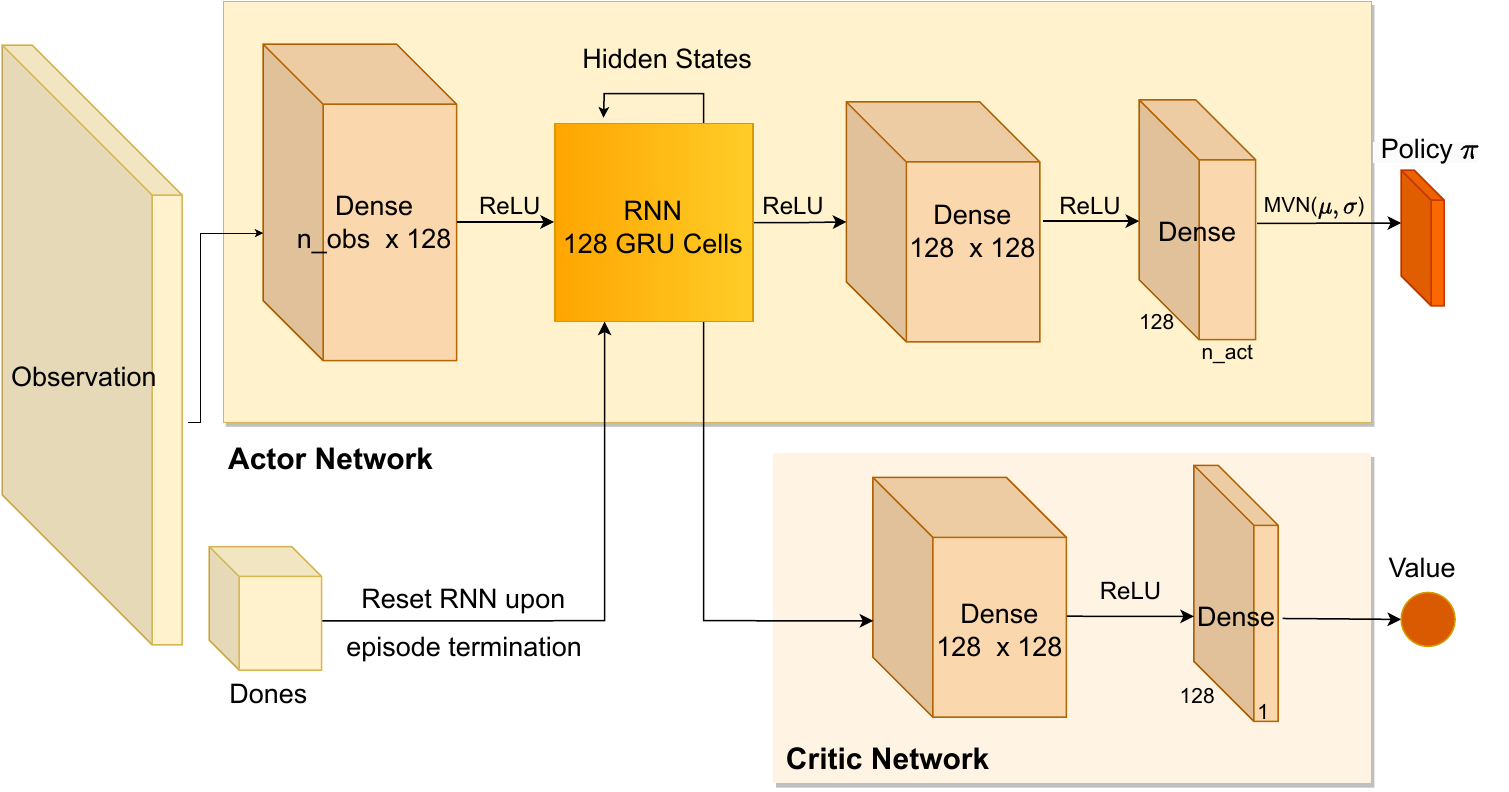}
  \caption{The architecture of the RNN-PPO actor and critic networks. The hidden states of the RNN layer add the necessary recurrence for the network to have memory of previous states. The Boolean termination values (``Dones'') are required to reset the memory of the RNN at episode termination. The policy $\pi$ is a multivariate normal distribution $mvn(\boldsymbol{\mu},\boldsymbol{\sigma})$ whereby $\boldsymbol{\sigma}$ is a trained network parameter and  $\boldsymbol{\mu}$ is the output vector from the final dense network layer.}
  \label{fig:architecture}
  \Description{}
  \vspace{-6mm}

\end{figure}

We re-implement Recurrent-PPO \citep{ma_opt_ex_ppo_fang22} to allow for a continuous action space. The full architecture of the actor and critic networks is given in Figure \ref{fig:architecture}. For every network update step, 10 steps are collected across 1,000 environments to create a batch size of 10,000 transitions. For each of the four epochs per update step, this batch is subdivided into four mini-batches which are used to calculate the loss functions, take the gradients and update the network through gradient descent using the Adam \citep{KingBa15} optimizer implemented in optax \citep{deepmind2020jax}.

\subsection{JAX enabled training speedup}
\label{sec:training_speedup}
\begin{figure}[h]
  \centering
  \vspace{-4mm}
  \includegraphics[width=0.95\linewidth]{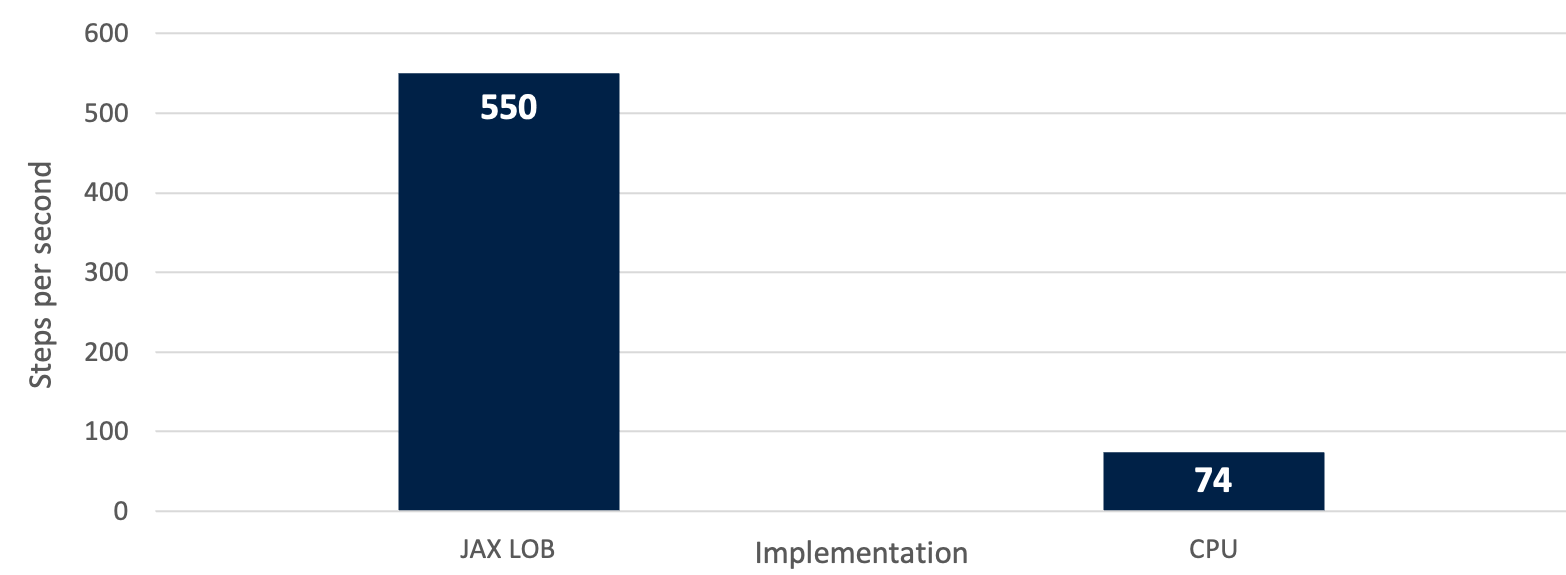}
  \caption{Training steps per second for gymnax environment based on JAX-LOB and our CPU gym environment implementation. The former is more than 7 times faster.}
  \label{fig:fps}

  \Description{}
  \vspace{-4mm}

\end{figure}

Whilst we do not have benchmarks for the exact same problem definition in other approaches, we nevertheless compare the training speeds for another environment running on the CPU. This is in addition to the reported data in Section \ref{sec:exec_costs} as the transfer of data from the CPU to GPU for the network update adds significant overhead to the run-time. With \textit{gymnax} running on the GPU, this is not necessary and we see further improvements in Figure \ref{fig:fps}. The RL4MM package, which we use for comparison in Sections \ref{sec:vmap} and \ref{sec:exec_costs}, is omitted in this comparison as we could not replicate fair experimental conditions due to issues arising with the use of the GPU. To qualify the comparison, we give a very brief overview of the CPU-based order book we use for the comparison. It is based on a LOB simulator using the traditional tree-like structure and uses the Stable-Baselines3 implementation of Recurrent PPO to solve an execution task. There are a number of key variables which are common across both experiments which we aim to control as far as possible to make the comparison equitable:
\begin{itemize}
    \item \textit{Messages processed per step} - 100 messages
    \item \textit{Levels of LOBSTER data considered} - 10 levels
    \item \textit{Episode length} - 30 minutes
    \item \textit{Hardware} - GPU: Nvidia A40, CPU: 32 core AMD EPYC 7513, both experiments run network updates on the GPU.
    \item \textit{Parallelization} - 1000 environments in parallel for JAX-LOB, 1000 vectorized environments for CPU but collected serially.
    \item \textit{Batch size} - 10,000, 10 steps across 1000 environments
    \item \textit{Mini-batch size} - 2,500, 4 per batch
    \item \textit{Number of Epochs} - 4 epochs
    \item \textit{Data} - April 2021 LOBSTER high frequency data for AMZN

\end{itemize}

We observe a 7x speedup of the training loop based on the JAX-LOB simulator compared to that using the CPU implementation of the LOB. This is larger than the reported 5x speedup in Table \ref{tab:matchvsquant}. We attribute this to the lack of data transfer across processing units. This does not yet consider hyper-parameter tuning which would further benefit from parallelization, and is trivial to achieve with our fully JAX-based implementation. 

\subsection{Training execution task}
\label{sec:training}
\begin{figure}[ht!]
\centering
\includegraphics[width=0.95\linewidth]{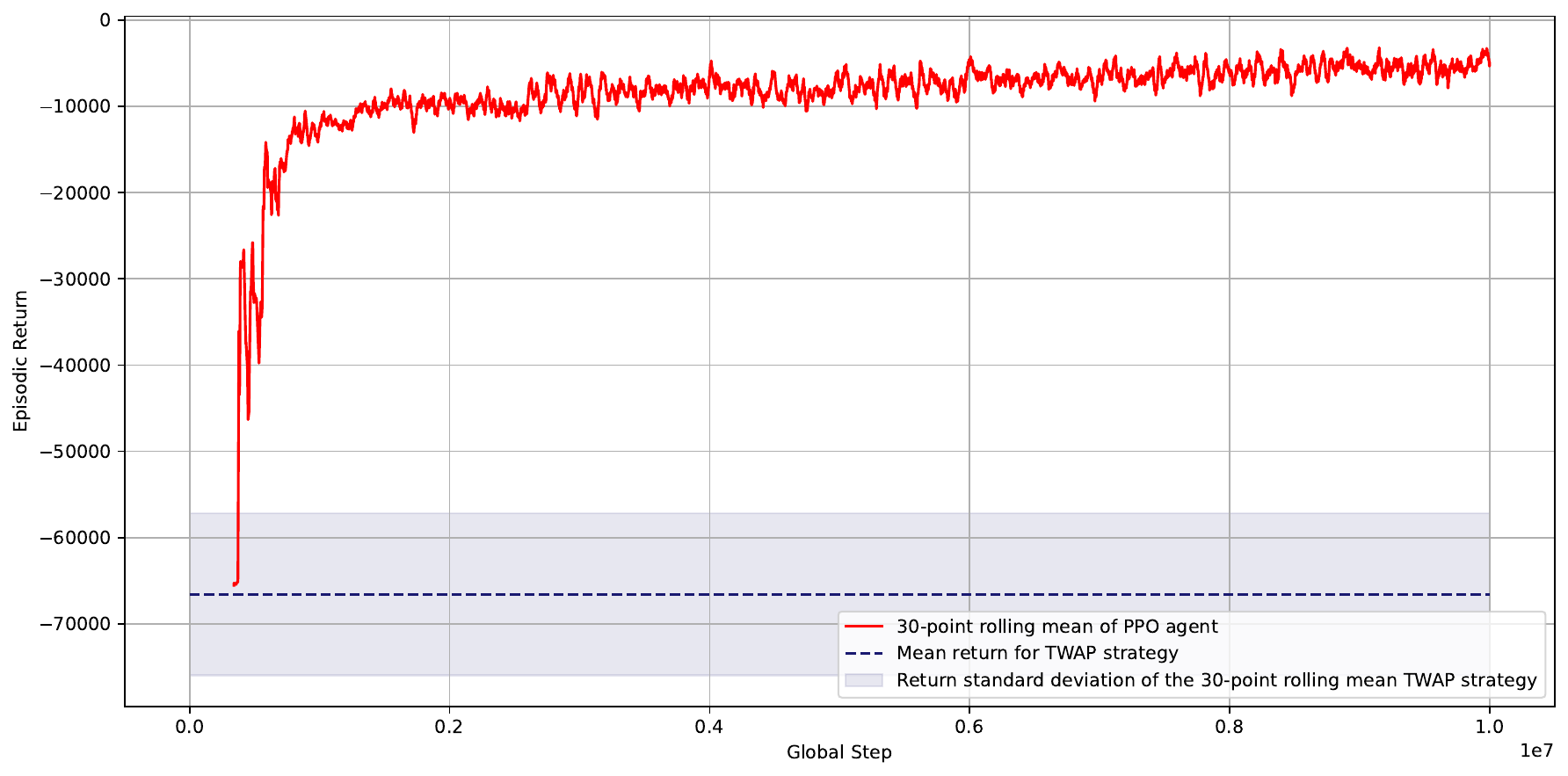}
\caption{Episodic return with $\lambda=0$ \eqref{eq:rewardfunc} for a PPO agent during training and the benchmark TWAP strategy mean return on a single day of data with 30-point rolling mean.}
\label{fig:training}
\vspace{-4mm}
\end{figure}

We begin to train an execution agent using the proposed setup on one day (April 1st, 2021) of data, chosen to be this narrow as the scope of this paper is primarily to showcase the functionality of the presented framework. The training curve, with a moving average window of 30 steps is shown in Figure \ref{fig:training}. We set $\lambda=0$ to make training easier, as the drift part of the reward function is much harder to learn, and compare the training curve to the TWAP benchmark strategy, which is to execute the desired volume linearly over the episode duration. Though the TWAP strategy is shown to be outperformed in training, we cannot make any claims on the success of this agent without conducting complete out-of-sample tests. Nevertheless, it indicates a promising direction of research based on the presented framework.

\section{Conclusions and future work}
\label{sec:conclusions}

We present the first implementation of a limit order book simulator for GPUs, called JAX-LOB. Using this as part of a \textit{gymnax} environment for reinforcement learning we obtain at least a 5x speedup compared to a comparable CPU implementation. When using the environment to train a RL agent with \textit{PureJaxRL} we even see a 7x speed-up with respect to our comparable CPU implementation. This speedup due to parallelization is expected to contribute to research in applying RL to high-frequency trading and execution problems that require a reactive LOB simulator.

As part of this paper, we provide an example use case of training an RL agent for the execution task on a single day of message data. We plan to extend this work by following a rigorous RL training pipeline for the execution problem amongst others, including out-of-sample testing and hyper-parameter tuning. The latter is expected to benefit further from the parallel nature of the JAX-LOB implementation. Detailed experiments with respect to the action-space, feature-space, reward shaping, and network architecture will be required to train a robust and cost-effective RL execution agent.

\bibliographystyle{ACM-Reference-Format}
\bibliography{zotero,software,execution,sample-base}


\begin{thebibliography}{36}


\ifx \showCODEN    \undefined \def \showCODEN     #1{\unskip}     \fi
\ifx \showDOI      \undefined \def \showDOI       #1{#1}\fi
\ifx \showISBNx    \undefined \def \showISBNx     #1{\unskip}     \fi
\ifx \showISBNxiii \undefined \def \showISBNxiii  #1{\unskip}     \fi
\ifx \showISSN     \undefined \def \showISSN      #1{\unskip}     \fi
\ifx \showLCCN     \undefined \def \showLCCN      #1{\unskip}     \fi
\ifx \shownote     \undefined \def \shownote      #1{#1}          \fi
\ifx \showarticletitle \undefined \def \showarticletitle #1{#1}   \fi
\ifx \showURL      \undefined \def \showURL       {\relax}        \fi
\providecommand\bibfield[2]{#2}
\providecommand\bibinfo[2]{#2}
\providecommand\natexlab[1]{#1}
\providecommand\showeprint[2][]{arXiv:#2}

\bibitem[Almgren and Chriss(2001)]%
        {almgren_optimal_2001}
\bibfield{author}{\bibinfo{person}{Robert Almgren} {and} \bibinfo{person}{Neil
  Chriss}.} \bibinfo{year}{2001}\natexlab{}.
\newblock \showarticletitle{Optimal execution of portfolio transactions}.
\newblock \bibinfo{journal}{\emph{The Journal of Risk}} \bibinfo{volume}{3},
  \bibinfo{number}{2} (\bibinfo{date}{Jan.} \bibinfo{year}{2001}),
  \bibinfo{pages}{5--39}.
\newblock
\urldef\tempurl%
\url{https://doi.org/10.21314/jor.2001.041}
\showDOI{\tempurl}
\newblock
\shownote{Publisher: Infopro Digital Services Limited}.


\bibitem[Amrouni et~al\mbox{.}(2022)]%
        {amrouni_abides-gym_2022}
\bibfield{author}{\bibinfo{person}{Selim Amrouni}, \bibinfo{person}{Aymeric
  Moulin}, \bibinfo{person}{Jared Vann}, \bibinfo{person}{Svitlana Vyetrenko},
  \bibinfo{person}{Tucker Balch}, {and} \bibinfo{person}{Manuela Veloso}.}
  \bibinfo{year}{2022}\natexlab{}.
\newblock \showarticletitle{{ABIDES}-gym: gym environments for multi-agent
  discrete event simulation and application to financial markets}. In
  \bibinfo{booktitle}{\emph{Proceedings of the {Second} {ACM} {International}
  {Conference} on {AI} in {Finance}}} \emph{(\bibinfo{series}{{ICAIF} '21})}.
  \bibinfo{publisher}{Association for Computing Machinery},
  \bibinfo{address}{New York, NY, USA}, \bibinfo{pages}{1--9}.
\newblock
\showISBNx{978-1-4503-9148-1}
\urldef\tempurl%
\url{https://doi.org/10.1145/3490354.3494433}
\showDOI{\tempurl}


\bibitem[Author(s)(2023)]%
        {nagy2023generative}
\bibfield{author}{\bibinfo{person}{Anonymous Author(s)}.}
  \bibinfo{year}{forthcoming 2023}\natexlab{}.
\newblock \showarticletitle{Generative AI for End-to-End Limit Order Book
  Modelling}.
\newblock  (\bibinfo{year}{forthcoming 2023}).
\newblock


\bibitem[Babuschkin et~al\mbox{.}(2020)]%
        {deepmind2020jax}
\bibfield{author}{\bibinfo{person}{Igor Babuschkin}, \bibinfo{person}{Kate
  Baumli}, \bibinfo{person}{Alison Bell}, \bibinfo{person}{Surya Bhupatiraju},
  \bibinfo{person}{Jake Bruce}, \bibinfo{person}{Peter Buchlovsky},
  \bibinfo{person}{David Budden}, \bibinfo{person}{Trevor Cai},
  \bibinfo{person}{Aidan Clark}, \bibinfo{person}{Ivo Danihelka},
  \bibinfo{person}{Antoine Dedieu}, \bibinfo{person}{Claudio Fantacci},
  \bibinfo{person}{Jonathan Godwin}, \bibinfo{person}{Chris Jones},
  \bibinfo{person}{Ross Hemsley}, \bibinfo{person}{Tom Hennigan},
  \bibinfo{person}{Matteo Hessel}, \bibinfo{person}{Shaobo Hou},
  \bibinfo{person}{Steven Kapturowski}, \bibinfo{person}{Thomas Keck},
  \bibinfo{person}{Iurii Kemaev}, \bibinfo{person}{Michael King},
  \bibinfo{person}{Markus Kunesch}, \bibinfo{person}{Lena Martens},
  \bibinfo{person}{Hamza Merzic}, \bibinfo{person}{Vladimir Mikulik},
  \bibinfo{person}{Tamara Norman}, \bibinfo{person}{George Papamakarios},
  \bibinfo{person}{John Quan}, \bibinfo{person}{Roman Ring},
  \bibinfo{person}{Francisco Ruiz}, \bibinfo{person}{Alvaro Sanchez},
  \bibinfo{person}{Rosalia Schneider}, \bibinfo{person}{Eren Sezener},
  \bibinfo{person}{Stephen Spencer}, \bibinfo{person}{Srivatsan Srinivasan},
  \bibinfo{person}{Wojciech Stokowiec}, \bibinfo{person}{Luyu Wang},
  \bibinfo{person}{Guangyao Zhou}, {and} \bibinfo{person}{Fabio Viola}.}
  \bibinfo{year}{2020}\natexlab{}.
\newblock \bibinfo{booktitle}{\emph{The {D}eep{M}ind {JAX} {E}cosystem}}.
\newblock
\urldef\tempurl%
\url{http://github.com/deepmind}
\showURL{%
\tempurl}


\bibitem[Belcak et~al\mbox{.}(2022)]%
        {belcak_fast_2022}
\bibfield{author}{\bibinfo{person}{Peter Belcak}, \bibinfo{person}{Jan-Peter
  Calliess}, {and} \bibinfo{person}{Stefan Zohren}.}
  \bibinfo{year}{2022}\natexlab{}.
\newblock \showarticletitle{Fast {Agent}-{Based} {Simulation} {Framework} with
  {Applications} to {Reinforcement} {Learning} and the {Study} of {Trading}
  {Latency} {Effects}}. In \bibinfo{booktitle}{\emph{Multi-{Agent}-{Based}
  {Simulation} {XXII}}} \emph{(\bibinfo{series}{Lecture {Notes} in {Computer}
  {Science}})}, \bibfield{editor}{\bibinfo{person}{Koen~H. Van~Dam} {and}
  \bibinfo{person}{Nicolas Verstaevel}} (Eds.). \bibinfo{publisher}{Springer
  International Publishing}, \bibinfo{address}{Cham}, \bibinfo{pages}{42--56}.
\newblock
\showISBNx{978-3-030-94548-0}
\urldef\tempurl%
\url{https://doi.org/10.1007/978-3-030-94548-0_4}
\showDOI{\tempurl}


\bibitem[Bertsimas and Lo(1998)]%
        {bertsimas_optimal_1998}
\bibfield{author}{\bibinfo{person}{Dimitris Bertsimas} {and}
  \bibinfo{person}{Andrew~W. Lo}.} \bibinfo{year}{1998}\natexlab{}.
\newblock \showarticletitle{Optimal control of execution costs}.
\newblock \bibinfo{journal}{\emph{Journal of Financial Markets}}
  \bibinfo{volume}{1}, \bibinfo{number}{1} (\bibinfo{date}{April}
  \bibinfo{year}{1998}), \bibinfo{pages}{1--50}.
\newblock
\showISSN{1386-4181}
\urldef\tempurl%
\url{https://doi.org/10.1016/S1386-4181(97)00012-8}
\showDOI{\tempurl}


\bibitem[Beysolow~II and Beysolow~II(2019)]%
        {beysolow2019market}
\bibfield{author}{\bibinfo{person}{Taweh Beysolow~II} {and}
  \bibinfo{person}{Taweh Beysolow~II}.} \bibinfo{year}{2019}\natexlab{}.
\newblock \showarticletitle{Market making via reinforcement learning}.
\newblock \bibinfo{journal}{\emph{Applied Reinforcement Learning with Python:
  With OpenAI Gym, Tensorflow, and Keras}} (\bibinfo{year}{2019}),
  \bibinfo{pages}{77--94}.
\newblock


\bibitem[Bouchaud et~al\mbox{.}(2018)]%
        {bouchaud_trades_2018}
\bibfield{author}{\bibinfo{person}{Jean-Philippe Bouchaud},
  \bibinfo{person}{Julius Bonart}, \bibinfo{person}{Jonathan Donier}, {and}
  \bibinfo{person}{Martin Gould}.} \bibinfo{year}{2018}\natexlab{}.
\newblock \bibinfo{booktitle}{\emph{Trades, {Quotes} and {Prices}: {Financial}
  {Markets} {Under} the {Microscope}}}.
\newblock \bibinfo{publisher}{Cambridge University Press}.
\newblock
\showISBNx{978-1-316-99888-5}
\newblock
\shownote{Google-Books-ID: dPRQDwAAQBAJ}.


\bibitem[Bradbury et~al\mbox{.}(2018)]%
        {jax2018github}
\bibfield{author}{\bibinfo{person}{James Bradbury}, \bibinfo{person}{Roy
  Frostig}, \bibinfo{person}{Peter Hawkins}, \bibinfo{person}{Matthew~James
  Johnson}, \bibinfo{person}{Chris Leary}, \bibinfo{person}{Dougal Maclaurin},
  \bibinfo{person}{George Necula}, \bibinfo{person}{Adam Paszke},
  \bibinfo{person}{Jake Vander{P}las}, \bibinfo{person}{Skye
  Wanderman-{M}ilne}, {and} \bibinfo{person}{Qiao Zhang}.}
  \bibinfo{year}{2018}\natexlab{}.
\newblock \bibinfo{booktitle}{\emph{{JAX}: composable transformations of
  {P}ython+{N}um{P}y programs}}.
\newblock
\urldef\tempurl%
\url{http://github.com/google/jax}
\showURL{%
\tempurl}


\bibitem[Byrd et~al\mbox{.}(2020)]%
        {byrd_abides_2020}
\bibfield{author}{\bibinfo{person}{David Byrd}, \bibinfo{person}{Maria
  Hybinette}, {and} \bibinfo{person}{Tucker~Hybinette Balch}.}
  \bibinfo{year}{2020}\natexlab{}.
\newblock \showarticletitle{{ABIDES}: {Towards} {High}-{Fidelity}
  {Multi}-{Agent} {Market} {Simulation}}. In
  \bibinfo{booktitle}{\emph{Proceedings of the 2020 {ACM} {SIGSIM} {Conference}
  on {Principles} of {Advanced} {Discrete} {Simulation}}}
  \emph{(\bibinfo{series}{{SIGSIM}-{PADS} '20})}.
  \bibinfo{publisher}{Association for Computing Machinery},
  \bibinfo{address}{New York, NY, USA}, \bibinfo{pages}{11--22}.
\newblock
\showISBNx{978-1-4503-7592-4}
\urldef\tempurl%
\url{https://doi.org/10.1145/3384441.3395986}
\showDOI{\tempurl}


\bibitem[Casgrain(2020)]%
        {julialob2020github}
\bibfield{author}{\bibinfo{person}{Philippe Casgrain}.}
  \bibinfo{year}{2020}\natexlab{}.
\newblock \bibinfo{booktitle}{\emph{LimitOrderBook.jl}}.
\newblock
\urldef\tempurl%
\url{https://github.com/p-casgrain/LimitOrderBook.jl}
\showURL{%
\tempurl}


\bibitem[Cont et~al\mbox{.}(2010)]%
        {cont_stochastic_2010}
\bibfield{author}{\bibinfo{person}{Rama Cont}, \bibinfo{person}{Sasha Stoikov},
  {and} \bibinfo{person}{Rishi Talreja}.} \bibinfo{year}{2010}\natexlab{}.
\newblock \showarticletitle{A {Stochastic} {Model} for {Order} {Book}
  {Dynamics}}.
\newblock \bibinfo{journal}{\emph{Operations Research}} \bibinfo{volume}{58},
  \bibinfo{number}{3} (\bibinfo{date}{June} \bibinfo{year}{2010}),
  \bibinfo{pages}{549--563}.
\newblock
\showISSN{0030-364X, 1526-5463}
\urldef\tempurl%
\url{https://doi.org/10.1287/opre.1090.0780}
\showDOI{\tempurl}


\bibitem[Dabérius et~al\mbox{.}(2019)]%
        {daberius19}
\bibfield{author}{\bibinfo{person}{Kevin Dabérius}, \bibinfo{person}{Elvin
  Granat}, {and} \bibinfo{person}{Patrik Karlsson}.}
  \bibinfo{year}{2019}\natexlab{}.
\newblock \showarticletitle{Deep execution-value and policy based reinforcement
  learning for trading and beating market benchmarks}.
\newblock \bibinfo{journal}{\emph{Available at SSRN 3374766}}
  (\bibinfo{year}{2019}).
\newblock


\bibitem[Fang et~al\mbox{.}(2022)]%
        {ma_opt_ex_ppo_fang22}
\bibfield{author}{\bibinfo{person}{Jin Fang}, \bibinfo{person}{Jiacheng Weng},
  \bibinfo{person}{Yi Xiang}, {and} \bibinfo{person}{Xinwen Zhang}.}
  \bibinfo{year}{2022}\natexlab{}.
\newblock \showarticletitle{Imitate then Transcend: Multi-Agent Optimal
  Execution with Dual-Window Denoise {PPO}}.
\newblock \bibinfo{journal}{\emph{arXiv preprint arXiv:2206.10736}}
  (\bibinfo{year}{2022}).
\newblock


\bibitem[Ganesh et~al\mbox{.}(2019)]%
        {ganesh2019reinforcement}
\bibfield{author}{\bibinfo{person}{Sumitra Ganesh}, \bibinfo{person}{Nelson
  Vadori}, \bibinfo{person}{Mengda Xu}, \bibinfo{person}{Hua Zheng},
  \bibinfo{person}{Prashant Reddy}, {and} \bibinfo{person}{Manuela Veloso}.}
  \bibinfo{year}{2019}\natexlab{}.
\newblock \showarticletitle{Reinforcement learning for market making in a
  multi-agent dealer market}.
\newblock \bibinfo{journal}{\emph{arXiv preprint arXiv:1911.05892}}
  (\bibinfo{year}{2019}).
\newblock


\bibitem[Google(23)]%
        {Google_7AD}
\bibfield{author}{\bibinfo{person}{Google}.} \bibinfo{year}{23}\natexlab{}.
\newblock \bibinfo{title}{XLA: Optimizing Compiler for machine learning\&nbsp;
  :\&nbsp; tensorflow}.
\newblock
\newblock
\urldef\tempurl%
\url{https://www.tensorflow.org/xla}
\showURL{%
\tempurl}


\bibitem[Gould et~al\mbox{.}(2013)]%
        {gould_limit_2013}
\bibfield{author}{\bibinfo{person}{Martin~D. Gould}, \bibinfo{person}{Mason~A.
  Porter}, \bibinfo{person}{Stacy Williams}, \bibinfo{person}{Mark McDonald},
  \bibinfo{person}{Daniel~J. Fenn}, {and} \bibinfo{person}{Sam~D. Howison}.}
  \bibinfo{year}{2013}\natexlab{}.
\newblock \showarticletitle{Limit order books}.
\newblock \bibinfo{journal}{\emph{Quantitative Finance}} \bibinfo{volume}{13},
  \bibinfo{number}{11} (\bibinfo{date}{Nov.} \bibinfo{year}{2013}),
  \bibinfo{pages}{1709--1742}.
\newblock
\showISSN{1469-7688}
\urldef\tempurl%
\url{https://doi.org/10.1080/14697688.2013.803148}
\showDOI{\tempurl}
\newblock
\shownote{Publisher: Routledge \_eprint:
  https://doi.org/10.1080/14697688.2013.803148}.


\bibitem[Huang and Polak(2011)]%
        {huang_lobster_2011}
\bibfield{author}{\bibinfo{person}{Ruihong Huang} {and} \bibinfo{person}{Tomas
  Polak}.} \bibinfo{year}{2011}\natexlab{}.
\newblock \bibinfo{title}{{LOBSTER}: {Limit} {Order} {Book} {Reconstruction}
  {System}}.
\newblock
\newblock
\urldef\tempurl%
\url{https://doi.org/10.2139/ssrn.1977207}
\showDOI{\tempurl}


\bibitem[Jerome et~al\mbox{.}(2022a)]%
        {jerome_market_2022}
\bibfield{author}{\bibinfo{person}{Joseph Jerome}, \bibinfo{person}{Gregory
  Palmer}, {and} \bibinfo{person}{Rahul Savani}.}
  \bibinfo{year}{2022}\natexlab{a}.
\newblock \showarticletitle{Market {Making} with {Scaled} {Beta} {Policies}}.
  In \bibinfo{booktitle}{\emph{Proceedings of the {Third} {ACM} {International}
  {Conference} on {AI} in {Finance}}}. \bibinfo{publisher}{ACM},
  \bibinfo{address}{New York NY USA}, \bibinfo{pages}{214--222}.
\newblock
\showISBNx{978-1-4503-9376-8}
\urldef\tempurl%
\url{https://doi.org/10.1145/3533271.3561745}
\showDOI{\tempurl}


\bibitem[Jerome et~al\mbox{.}(2022b)]%
        {jerome_model-based_2022}
\bibfield{author}{\bibinfo{person}{Joseph Jerome}, \bibinfo{person}{Leandro
  Sanchez-Betancourt}, \bibinfo{person}{Rahul Savani}, {and}
  \bibinfo{person}{Martin Herdegen}.} \bibinfo{year}{2022}\natexlab{b}.
\newblock \bibinfo{title}{Model-based gym environments for limit order book
  trading}.
\newblock
\newblock
\urldef\tempurl%
\url{http://arxiv.org/abs/2209.07823}
\showURL{%
\tempurl}
\newblock
\shownote{arXiv:2209.07823 [cs, q-fin]}.


\bibitem[Karpe et~al\mbox{.}(2020)]%
        {karpe20}
\bibfield{author}{\bibinfo{person}{Michäel Karpe}, \bibinfo{person}{Jin Fang},
  \bibinfo{person}{Zhongyao Ma}, {and} \bibinfo{person}{Chen Wang}.}
  \bibinfo{year}{2020}\natexlab{}.
\newblock \showarticletitle{Multi-agent reinforcement learning in a realistic
  limit order book market simulation}. In \bibinfo{booktitle}{\emph{Proceedings
  of the First ACM International Conference on AI in Finance}}.
  \bibinfo{publisher}{ACM}.
\newblock


\bibitem[Kingma and Ba(2015)]%
        {KingBa15}
\bibfield{author}{\bibinfo{person}{Diederik Kingma} {and}
  \bibinfo{person}{Jimmy Ba}.} \bibinfo{year}{2015}\natexlab{}.
\newblock \showarticletitle{Adam: A Method for Stochastic Optimization}. In
  \bibinfo{booktitle}{\emph{International Conference on Learning
  Representations (ICLR)}}. \bibinfo{address}{San Diega, CA, USA}.
\newblock


\bibitem[Kochkov et~al\mbox{.}(2021)]%
        {kochkov_machine_2021}
\bibfield{author}{\bibinfo{person}{Dmitrii Kochkov}, \bibinfo{person}{Jamie~A.
  Smith}, \bibinfo{person}{Ayya Alieva}, \bibinfo{person}{Qing Wang},
  \bibinfo{person}{Michael~P. Brenner}, {and} \bibinfo{person}{Stephan Hoyer}.}
  \bibinfo{year}{2021}\natexlab{}.
\newblock \showarticletitle{Machine learning–accelerated computational fluid
  dynamics}.
\newblock \bibinfo{journal}{\emph{Proceedings of the National Academy of
  Sciences}} \bibinfo{volume}{118}, \bibinfo{number}{21} (\bibinfo{date}{May}
  \bibinfo{year}{2021}), \bibinfo{pages}{e2101784118}.
\newblock
\urldef\tempurl%
\url{https://doi.org/10.1073/pnas.2101784118}
\showDOI{\tempurl}
\newblock
\shownote{Publisher: Proceedings of the National Academy of Sciences}.


\bibitem[Kyle(1985)]%
        {kyle_continuous_1985}
\bibfield{author}{\bibinfo{person}{Albert~S. Kyle}.}
  \bibinfo{year}{1985}\natexlab{}.
\newblock \showarticletitle{Continuous {Auctions} and {Insider} {Trading}}.
\newblock \bibinfo{journal}{\emph{Econometrica}} \bibinfo{volume}{53},
  \bibinfo{number}{6} (\bibinfo{year}{1985}), \bibinfo{pages}{1315--1335}.
\newblock
\showISSN{0012-9682}
\urldef\tempurl%
\url{https://doi.org/10.2307/1913210}
\showDOI{\tempurl}
\newblock
\shownote{Publisher: [Wiley, Econometric Society]}.


\bibitem[Lange(2022)]%
        {gymnax2022github}
\bibfield{author}{\bibinfo{person}{Robert~Tjarko Lange}.}
  \bibinfo{year}{2022}\natexlab{}.
\newblock \bibinfo{booktitle}{\emph{{gymnax}: A {JAX}-based Reinforcement
  Learning Environment Library}}.
\newblock
\urldef\tempurl%
\url{http://github.com/RobertTLange/gymnax}
\showURL{%
\tempurl}


\bibitem[Lu et~al\mbox{.}(2022)]%
        {lu2022discovered}
\bibfield{author}{\bibinfo{person}{Chris Lu}, \bibinfo{person}{Jakub Kuba},
  \bibinfo{person}{Alistair Letcher}, \bibinfo{person}{Luke Metz},
  \bibinfo{person}{Christian Schroeder~de Witt}, {and} \bibinfo{person}{Jakob
  Foerster}.} \bibinfo{year}{2022}\natexlab{}.
\newblock \showarticletitle{Discovered policy optimisation}.
\newblock \bibinfo{journal}{\emph{Advances in Neural Information Processing
  Systems}}  \bibinfo{volume}{35} (\bibinfo{year}{2022}),
  \bibinfo{pages}{16455--16468}.
\newblock


\bibitem[Nevmyvaka et~al\mbox{.}(2006)]%
        {nevmyvaka2006reinforcement}
\bibfield{author}{\bibinfo{person}{Yuriy Nevmyvaka}, \bibinfo{person}{Yi Feng},
  {and} \bibinfo{person}{Michael Kearns}.} \bibinfo{year}{2006}\natexlab{}.
\newblock \showarticletitle{Reinforcement learning for optimized trade
  execution}. In \bibinfo{booktitle}{\emph{Proceedings of the 23rd
  international conference on Machine learning}}. \bibinfo{pages}{673--680}.
\newblock


\bibitem[Ning et~al\mbox{.}(2021)]%
        {ddqn_opt_ex_ning18}
\bibfield{author}{\bibinfo{person}{Brian Ning}, \bibinfo{person}{Franco Ho~Ting
  Lin}, {and} \bibinfo{person}{Sebastian Jaimungal}.}
  \bibinfo{year}{2021}\natexlab{}.
\newblock \showarticletitle{Double deep {Q}-learning for optimal execution}.
\newblock \bibinfo{journal}{\emph{Applied Mathematical Finance}}
  \bibinfo{volume}{28}, \bibinfo{number}{4} (\bibinfo{year}{2021}),
  \bibinfo{pages}{361--380}.
\newblock


\bibitem[Obizhaeva and Wang(2013)]%
        {obizhaeva_optimal_2013}
\bibfield{author}{\bibinfo{person}{Anna~A. Obizhaeva} {and}
  \bibinfo{person}{Jiang Wang}.} \bibinfo{year}{2013}\natexlab{}.
\newblock \showarticletitle{Optimal trading strategy and supply/demand
  dynamics}.
\newblock \bibinfo{journal}{\emph{Journal of Financial Markets}}
  \bibinfo{volume}{16}, \bibinfo{number}{1} (\bibinfo{date}{Feb.}
  \bibinfo{year}{2013}), \bibinfo{pages}{1--32}.
\newblock
\showISSN{1386-4181}
\urldef\tempurl%
\url{https://doi.org/10.1016/j.finmar.2012.09.001}
\showDOI{\tempurl}


\bibitem[Paulin(2019)]%
        {paulin_understanding_2019_custom}
\bibfield{author}{\bibinfo{person}{James Paulin}.}
  \bibinfo{year}{2019}\natexlab{}.
\newblock \bibinfo{title}{Understanding flash crash contagion and systemic
  risk: a calibrated agent-based approach}.
\newblock
\newblock
\urldef\tempurl%
\url{https://ora.ox.ac.uk/objects/uuid:929fa3fe-4e5f-4cef-ad9f-03eb40110818}
\showURL{%
\tempurl}


\bibitem[Quera-Bofarull et~al\mbox{.}(2023)]%
        {quera-bofarull_challenges_2023}
\bibfield{author}{\bibinfo{person}{Arnau Quera-Bofarull}, \bibinfo{person}{Joel
  Dyer}, \bibinfo{person}{Anisoara Calinescu}, {and} \bibinfo{person}{Michael
  Wooldridge}.} \bibinfo{year}{2023}\natexlab{}.
\newblock \bibinfo{title}{Some challenges of calibrating differentiable
  agent-based models}.
\newblock
\newblock
\urldef\tempurl%
\url{http://arxiv.org/abs/2307.01085}
\showURL{%
\tempurl}
\newblock
\shownote{arXiv:2307.01085 [cs, q-fin, stat]}.


\bibitem[Schoenholz and Cubuk(2020)]%
        {jaxmd2020}
\bibfield{author}{\bibinfo{person}{Samuel~S. Schoenholz} {and}
  \bibinfo{person}{Ekin~D. Cubuk}.} \bibinfo{year}{2020}\natexlab{}.
\newblock \showarticletitle{JAX M.D. A Framework for Differentiable Physics}.
  In \bibinfo{booktitle}{\emph{Advances in Neural Information Processing
  Systems}}, Vol.~\bibinfo{volume}{33}. \bibinfo{publisher}{Curran Associates,
  Inc.}
\newblock
\urldef\tempurl%
\url{https://papers.nips.cc/paper/2020/file/83d3d4b6c9579515e1679aca8cbc8033-Paper.pdf}
\showURL{%
\tempurl}


\bibitem[Scholl et~al\mbox{.}(2021)]%
        {scholl_how_2021}
\bibfield{author}{\bibinfo{person}{Maarten~P. Scholl},
  \bibinfo{person}{Anisoara Calinescu}, {and} \bibinfo{person}{J.~Doyne
  Farmer}.} \bibinfo{year}{2021}\natexlab{}.
\newblock \showarticletitle{How market ecology explains market malfunction}.
\newblock \bibinfo{journal}{\emph{Proceedings of the National Academy of
  Sciences}} \bibinfo{volume}{118}, \bibinfo{number}{26} (\bibinfo{date}{June}
  \bibinfo{year}{2021}), \bibinfo{pages}{e2015574118}.
\newblock
\urldef\tempurl%
\url{https://doi.org/10.1073/pnas.2015574118}
\showDOI{\tempurl}
\newblock
\shownote{Publisher: Proceedings of the National Academy of Sciences}.


\bibitem[Schulman et~al\mbox{.}(2017)]%
        {ppo_schulman17}
\bibfield{author}{\bibinfo{person}{John Schulman}, \bibinfo{person}{Filip
  Wolski}, \bibinfo{person}{Prafulla Dhariwal}, \bibinfo{person}{Alec Radford},
  {and} \bibinfo{person}{Oleg Klimov}.} \bibinfo{year}{2017}\natexlab{}.
\newblock \showarticletitle{Proximal Policy Optimization Algorithms}.
\newblock \bibinfo{journal}{\emph{arXiv preprint arXiv:1707.06347}}
  (\bibinfo{year}{2017}).
\newblock


\bibitem[Van~Hasselt et~al\mbox{.}(2016)]%
        {dddqn_hasselt2016}
\bibfield{author}{\bibinfo{person}{Hado Van~Hasselt}, \bibinfo{person}{Arthur
  Guez}, {and} \bibinfo{person}{David Silver}.}
  \bibinfo{year}{2016}\natexlab{}.
\newblock \showarticletitle{Deep reinforcement learning with double
  {Q}-learning}. In \bibinfo{booktitle}{\emph{Proceedings of the AAAI
  conference on artificial intelligence}}, Vol.~\bibinfo{volume}{30}.
\newblock


\bibitem[Vyetrenko et~al\mbox{.}(2020)]%
        {vyetrenko_get_2020}
\bibfield{author}{\bibinfo{person}{Svitlana Vyetrenko}, \bibinfo{person}{David
  Byrd}, \bibinfo{person}{Nick Petosa}, \bibinfo{person}{Mahmoud Mahfouz},
  \bibinfo{person}{Danial Dervovic}, \bibinfo{person}{Manuela Veloso}, {and}
  \bibinfo{person}{Tucker Balch}.} \bibinfo{year}{2020}\natexlab{}.
\newblock \showarticletitle{Get real: realism metrics for robust limit order
  book market simulations}. In \bibinfo{booktitle}{\emph{Proceedings of the
  {First} {ACM} {International} {Conference} on {AI} in {Finance}}}
  \emph{(\bibinfo{series}{{ICAIF} '20})}. \bibinfo{publisher}{Association for
  Computing Machinery}, \bibinfo{address}{New York, NY, USA},
  \bibinfo{pages}{1--8}.
\newblock
\showISBNx{978-1-4503-7584-9}
\urldef\tempurl%
\url{https://doi.org/10.1145/3383455.3422561}
\showDOI{\tempurl}


\end{thebibliography}

\appendix

\end{document}